\documentclass{mpe_report}
\usepackage{psfig}

\begin{document}
\title{``X--ray supernova remnants in nearby galaxies''}
\author{I. Leonidaki\inst{1,3} \,A. Zezas\inst{2} \,P. Boumis\inst{1}}  
\institute{Institute of Astronomy \& Astrophysics, National Observatory of Athens, Athens, Greece 
\and  Harvard -- Smithsonian Center for Astrophysics, Cambridge, MA, USA
\and  Department of Physics, University of Patras, Rio--Patras, Greece}
\maketitle

\begin{abstract}
We present the initial results from a study of the SNR population in a
sample of six nearby galaxies (NGC 2403, NGC 4214, NGC 4449, NGC 5204,
NGC 3077, NGC 4395) based on Chandra archival data. We discuss the analysis of
the Chandra data and we present candidate SNR sources selected on the
basis of their X--ray colours. We also present deep [S {\sc ii}] 6716 \&
6731 ${\rm \AA}$~and H$\alpha$ line images for most of the galaxies in our
sample, which provide optically selected samples of SNRs. Comparison
of the X--ray results with the complementary optical observations
provides a more complete picture of the SNR population and allows us
to address their X--ray emission. Our preliminary analysis of the [S
{\sc ii}]/H$\alpha$ images show that 48 X--ray sources are typically
associated with H$\alpha$ sources, 7 of which are SNR candidates based
on their [S {\sc ii}]/H$\alpha$ ratio and one is an already known
radio SNR.
\end{abstract}

\begin{figure}
\centerline{\psfig{file=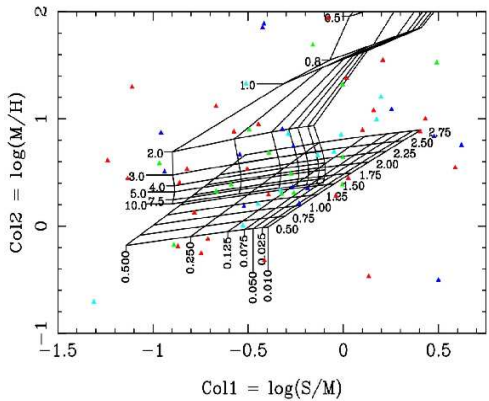,width=6.5cm,clip=} }
\caption{X--ray Color--color diagnostic diagram.
\label{image}}
\end{figure}

\section{Introduction}

Supernova Remnants (SNRs) are an important component of the X--ray
source populations in galaxies (Blair \& Long 1997), especially in
luminosities below $10^{37}$ erg sec$^ {-1}$. The expanding SNR
shocks heat the surrounding interstellar medium (ISM) to temperatures
of $10^6$ -- $10^7$ K and produce X--ray emission. This way SNRs
provide a significant fraction of the mechanical energy that heats,
shapes and chemically enriches the ISM. Despite extensive studies of
individual galactic SNRs, a complete understanding of these sources
remains elusive. It is well known that different wavebands provide a
picture of different evolutionary stages of the SNR populations. On
the other hand, although it is almost certain that environment plays a
major role in the evolution and the multiwavelength properties of SNRs
(e.g. Pannuti et al. 2007), the details of this connection are poorly
understood. In order to address this connection, we initiated an X--ray
and optical study of six nearby galaxies. Here we present the
preliminary results from this work focusing on the identification and
classification of the SNR candidates.

\begin{figure}
\centerline{\psfig{file=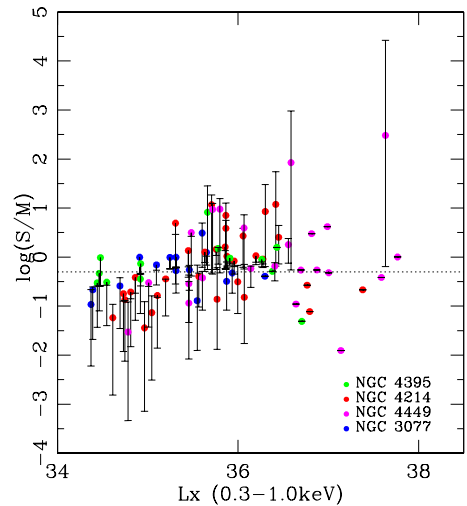,width=7.7cm,clip=} }
\caption{X-ray color--luminosity plot.
\label{image}}
\end{figure}

\section{Sample, data reduction and identification method}

Our sample consists of the nearby galaxies observed with Chandra and
fulfil the following criteria : a) Distance $>$ 5 Mpc, in order to
minimize source confusion; b) Inclination $>$ 60 degrees, in order
to minimize internal extinction and projection effects; and c)
Exposure times long enough to achieve a uniform detection limit of
~$10^{36}$ erg sec$^ {-1}$. The X--ray data analysis was performed with the CIAO
tool suite version 3.4 and with custom developed scripts. This
analysis includes initial cleaning of the data, source detection,
photometry and multi band image extraction. The identification of the
thermal SNRs is based on their hardness ratios (e.g. Prestwich et
al. 2003). X--ray colours can be used to obtain information on the
spectral properties of X--ray sources, especially in the case of small
number of counts when spectral fitting is not possible.  We calculate
the X--ray colours using the BEHR tool (Park et al. 2006). In addition
to the Chandra data, we obtained optical images in the [S {\sc ii}]
6716 \& 6731 ${\rm \AA}$ and H$\alpha$ lines using the 1.3m Skinakas
(Greece) and the 1.2m FLWO (USA) telescopes. The optical images were
also cleaned, sky subtracted and aligned using the IRAF package. The
final H$\alpha$ and [S {\sc ii}] images were divided in order to
locate sources with ratio [S {\sc ii}]/H$\alpha$ $>$ 0.4
(Mathewson \& Clarke 1973), indicating optical SNR candidates.

\begin{figure}
\centerline{\psfig{file=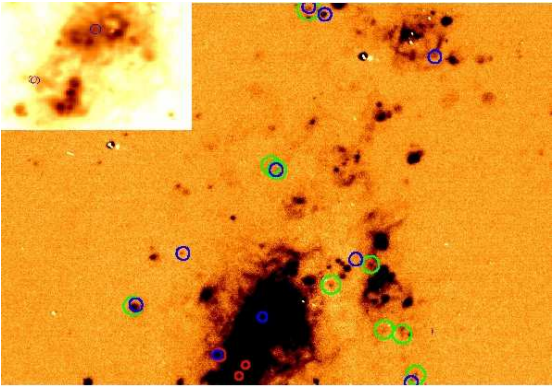,width=7.3cm,clip=} }
\caption{H$\alpha$ image of NGC 4214. Upper left image: zoomed,
rescaled and with different contrast image of the region with the
known radio SNRs.  \label{image}} \end{figure}

\section{Preliminary Results}

In Figure 1 we show a plot of the X--ray color Col1=log(S/M) versus
the X--ray color Col2=log(M/H), where S, M, H are the numbers of
counts in the soft (0.1 -- 3.0 keV), medium (1.0 -- 2.5 keV) and hard band
(2.5 -- 7.0 keV), of all the detected sources in four galaxies of our
sample. On the same plot we present X--ray grids for power law and
thermal plasma models for different temperature (kT), absorbing H{\sc
i} column density ${\rm N_{H}}$ and photon index ($\Gamma$). The circle shows
the locus of thermal SNRs given that they are expected to have
temperatures of $<$ 2 keV.  From the same diagram we see that sources
with Col1 $>$ --0.3 have a high proportion of candidate SNRs. The grids
are calculated based on the average affective area of the
observations. The points correspond to the hardness ratios for each
source calculated using the BEHR method (for clarity we do not show
error bars). Points outside the grids have large errors and/or complex
multi--component models. The red circle shows the locus of SNRs used
in our analysis.  Figure 2 shows a plot of Col1 versus luminosity.The
X--ray SNR candidates are those with Col1 $>$ --0.3. We see that a large
number of these sources are soft with luminosities around
$\sim$$10^{36}$ erg s$^ {-1}$ cm$^{-2}$. Based on
the color -- color diagram, many of the sources can be absorbed SNRs.


\begin{acknowledgements}

This work has been partly supported by NASA grant GO6-7086X and NASA
LTSA grant G5-13056. IL wishes to thank the Center for Astrophysics
for its hospitality during her visits there and acknowledges funding
by the European Union and the Greek Ministry of Development in the
framework of the programme 'Promotion of Excellence in Research
Institutes (2nd Part)' and the European Social Fund (ESF), Operational
Program for Educational and Vocational Training II (EPEAEK II), and
particularly the Program PYTHAGORAS II. The authors would like to
thank the staff at 1.3 Skinakas Observatory (Crete, Greece) and FLWO
1.2 Observatory (Mt. Hopkins, Arizona, USA) for their excellent
support during the observations. Skinakas Observatory is a
collaborative project of the University of Crete, the Foundation for
Research and the Technology-Hellas and the Max-Planck-Institute.

\end{acknowledgements}

\begin{figure}
\centerline{\psfig{file=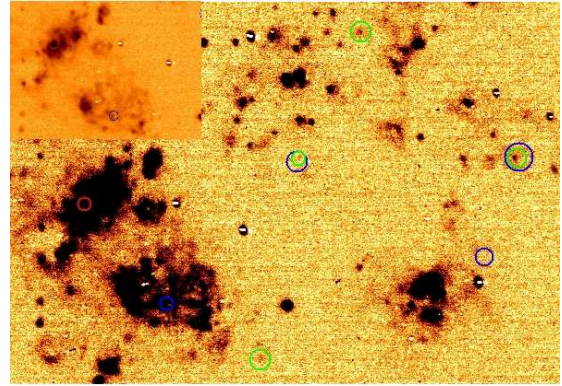,width=7.3cm,clip=} }
\caption{H$\alpha$ image of NGC 4395. Upper left image: zoomed,
rescaled and with different contrast image of the region with the
known radio SNRs. \label{image}}
\end{figure}
   


\begin{thebibliography}{} 
\bibitem{BL97} Blair, W.P. \& Long, K.S., 1997,ApJS, 108, 261
\bibitem{MC73} Mathewson, D.S., \& Clarke, L.N., 1973, ApJ, 180, 725 
\bibitem{MF97} Mattonick, D.M. \& Fesen, R.A., 1997, ApJS, 112, 49
\bibitem{M97}Mattonick, D.M. et al., 1997, ApJS, 113, 333
\bibitem{PSG07} Panutti, T., Schlegel, E.M., Griffith, S.A., 2007, AJ, 210, 1510
\bibitem{P06} Park, T. et al., 2006, ApJ, 652, 610
\bibitem{P04} Prestwich, A., et al., 2004, A\&A, 595, 719
\bibitem{V05} Vukotic, 2005, SerAJ, 170, 101
\end{thebibliography}
\end{document}